\documentclass[pra,preprint,showpacs,showkeys,amsfonts]{revtex4}
\usepackage{graphicx}
\RequirePackage{times}
\RequirePackage{mathptm}
\begin{document}

\title{Feyerabend and physics\footnote{Presented at the International Symposium Paul Feyerabend 1924-1994. A philosopher from Vienna, University of Vienna, June 18-19, 2004}}
\author{Karl Svozil}
 \email{svozil@tuwien.ac.at}
\homepage{http://tph.tuwien.ac.at/~svozil}
\affiliation{Institut f\"ur Theoretische Physik, University of Technology Vienna,
Wiedner Hauptstra\ss e 8-10/136, A-1040 Vienna, Austria}

\begin{abstract}
Feyerabend frequently discussed physics.
He also referred to the history of the subject when motivating his philosophy of science.
Alas, as some examples show, his understanding of physics remained superficial.
In this respect, Feyerabend is like Popper;
the difference being his self-criticism later on,
and the much more tolerant attitude toward the allowance of methods.
Quite generally, partly due to the complexity of the formalism and the new challenges of their findings,
which left philosophy proper at a loss,
physicists have attempted to developed their own
meaning of their subject.
For instance, in recent years, the interpretation of quantum mechanics
has stimulated a new type of experimental philosophy,
which seeks to operationalize emerging philosophical issues;
issues which are incomprehensible for most philosophers.
In this respect, physics often appears to be a continuation of philosophy by other means.
Yet, Feyerabend has also expressed profound insights into the possibilities for the progress of physics,
a legacy which remains to be implemented in the times to come:
the conquest of abundance, the richness of reality,
the many worlds which still await discovery,
and the vast openness of the physical universe.
\end{abstract}


\maketitle

\section{General attitude}

Starting with his dissertation at the University of Vienna under Hans Thirring and Viktor Kraft,
Feyerabend wrote some papers on
physics-related topics, in particular on the interpretation on quantum mechanics,
on classical  and  statistical physics.
In his autobiography he himself admitted to be no real expert in this area,
noting that he had started calculating a problem of classical electrodynamics but
seemed to be getting nowhere (p.~85 of Ref.~\cite{feyerabend-auto}).
Unlike Popper's attempts to ``falsify the Copenhagen interpretation''
\cite{2002-peres} and argue against the quantum logic introduced by Birkhoff and von Neumann
\cite{dalla-2002},
Feyerabend pursued these investigations in a more humble, considerate  and  self-critcal style.

Professor Fischer recalls \cite{fischer} that the physicists at Berkeley
generously evaluated Feyerabend to be two decades behind current research,
the average philosopher being at least half-a-century behind.
Also, Fischer recalls, Feyerabend was happy with the evaluation, and told him
that there was no essential difference between a physicist and a good philosopher,
and that Feyerabend considered himself to be too stupid to make discoveries in physics to be
considered physicist: {\em ``Apart from his stupidity --- he assured me --- nothing separated him from
being a physicist.''}

In contrast to formalized physics, Feyerabend's contributions and insights into methodological issues are,
at least in my opinion, as remarkable as they are provocative;
sometimes even bordering to the offensive; always gathering attention and raising eyebrows.
Sometimes  the reaction were harsh.
In an article published in {\em Nature}
(p.~596 of Ref.~\cite{theo-psi-1987}), Feyerabend was referred to as
{\em ``the Salvador Dali of academic philosophy, and currently the worst enemy of science;''}
a denunciation which deeply saddened him (Chapter 12 of Ref.~\cite{feyerabend-auto}).
I do not think that such a term is justified.
Popper with his naive viewpoint and his talk about {\em ``blablabla''}
certainly did more harm to science \cite{svozil-2002-popper}
than any other dillettante claiming to know the proceeds of science before;
but not Feyerabend.
 On the contrary I believe that Feyerabend was right in suggesting
that input from the outside does science proper good;
sometimes this seems even mandatory, even if one is not willing to grant that
{\em ``science has now become as oppressive as the ideologies it had once to fight''}
\cite{feyerabend-defense,feyer-81}.

Besides his methodological openness,
Feyerabend's lasting message, in my opinion, is the ``conquest of abundance,''
the ``richness'' of the phenomena concerning us, and
the ``vasteness'' of the territories still awaiting to be discovered.
The pursuit of science is one of the greatest passions of life,
and our capabilities to recognize and manipulate the physical world
may only be limited by our phantasy.
Maybe one hopefully happy day we will be able to {\em tune} the world according to our will alone.

\section{Tower of Pisa example in ``Against Method''}

One of the things which Feyerabend discussed in {\em Against Method} \cite{feyerabend} in greater detail
is the Tower of Pisa example.
It is about an old argument against earth rotation which has been already put forward by Aristotle:
A stone from a high tower arrives at the foot of the tower without
any shift relative to the horizontal position of the release point on top of the tower.

Admittedly, Feyerabend had other objectives in mind,
in particular some supposed   ``deceptions'' by Galileo,
who allegedly ``brushed aside'' topics seemingly in conflict with his heliocentric approach by maintaining that the phenomena
could be correctly described while at the same time
``hiding'' new ``absurd'' theoretical assumptions.

Indeed, Galileo seems to have committed himself to the attitude that  there should be no shift whatsoever,
a conjecture which also seemed to have been accepted  by Copernicus.
Newton and Hook investigated this topic more carefully.
Indeed, this may have been the starting point of Newton's theory of gravity.
Also Gauss  and  Laplace held (wrong) theoretical opinions on the phenomenon.

After a succession of inconclusive measurements by different researchers,
Hall performed  experiments in Harvard in 1902
 \cite{hall-1903a,hall-1903b}.
Due to the admirable effort of the {\em American Physical Society}
to retroscan their entire collection of scholarly articles published in the {\em Physical Reviews},
Hall's superbly written contributions are easily obtainable.
A later review by Armitage \cite{armitage} which is also cited in  {\em Against Method}
states,
{\em ``$\ldots$ Thus Newton's experimental test for the diurnal rotation
of the Earth may be said to have given positive results of the expected
order of magnitude, though the persistent occurrence of an unaccountable southward deviation
has continued to be a matter for inconclusive speculation.''}

Despite our present conception of a ferocious earth rotation,
which reaches its peak of  464 m/sec or  1670 km/hour at the equator,
and which may give rise to measurable effects even if the relative motions are assumed to be small,
in his writings Feyerabend never mentioned the contemporary physical situation,
in particular the Coriolis force  and  the Kepler problem.
This seems to be characteristic for the attitude of many philosophers of science,
as Feyerabend himself polemically notes \cite{feyerabend-defense,feyer-81},
{\em ``$\ldots$ Kuhn encourages people who have no idea why a stone falls to the ground
to talk with assurance about scientific method.
Now I have no objection to incompetence but I do object when incompetence is accompanied by boredom
and self-righteousness.
And this is exactly what happens. $\ldots$''}
When one reads these strong words,
written in an intellectual climate of the seventies of the past century,
one has little doubt that the boldness and self-esteem of such statements
provoked antagonism.

Coming back to Tower of Pisa example,
some model calculation were done by Martina Jedinger  and   Iva Brezinova here in Vienna,
yielding a latitudinal shift of 9.6 cm towards South and a longitudinal shift of
0.6 cm towards East. Intuitively, the large latitudinal shift could be understood
by considering that (air resistance left aside), the falling body remains in a plain
spanned by the direction of the gravity pull towards the center of the earth,
and by the direction of velocity at its release point.
At the same time, the earth, and with it the foot of the tower, revolves around an axis
which is currently tilted at 23.5$^{\circ}$ with respect to the ecliptic axis,
the line drawn from the center of the earth and perpendicular to the ecliptic plane;
a configuration depicted in  Fig.~\ref{f-2004-f1}.
\begin{figure}
\begin{center}
\unitlength 0.30mm
\linethickness{0.4pt}
\begin{picture}(100.00,111.67)
\multiput(50.00,100.00)(1.60,-0.10){4}{\line(1,0){1.60}}
\multiput(56.39,99.59)(0.57,-0.11){11}{\line(1,0){0.57}}
\multiput(62.68,98.36)(0.36,-0.12){17}{\line(1,0){0.36}}
\multiput(68.76,96.35)(0.24,-0.12){24}{\line(1,0){0.24}}
\multiput(74.54,93.57)(0.18,-0.12){30}{\line(1,0){0.18}}
\multiput(79.91,90.07)(0.14,-0.12){35}{\line(1,0){0.14}}
\multiput(84.78,85.92)(0.12,-0.13){36}{\line(0,-1){0.13}}
\multiput(89.09,81.17)(0.12,-0.17){31}{\line(0,-1){0.17}}
\multiput(92.76,75.92)(0.12,-0.23){25}{\line(0,-1){0.23}}
\multiput(95.72,70.24)(0.12,-0.32){19}{\line(0,-1){0.32}}
\multiput(97.93,64.23)(0.12,-0.52){12}{\line(0,-1){0.52}}
\multiput(99.36,57.98)(0.10,-1.06){6}{\line(0,-1){1.06}}
\multiput(99.97,51.60)(-0.10,-3.20){2}{\line(0,-1){3.20}}
\multiput(99.77,45.20)(-0.11,-0.70){9}{\line(0,-1){0.70}}
\multiput(98.75,38.87)(-0.11,-0.38){16}{\line(0,-1){0.38}}
\multiput(96.92,32.73)(-0.12,-0.27){22}{\line(0,-1){0.27}}
\multiput(94.33,26.87)(-0.12,-0.20){28}{\line(0,-1){0.20}}
\multiput(91.01,21.39)(-0.12,-0.15){34}{\line(0,-1){0.15}}
\multiput(87.01,16.38)(-0.12,-0.12){38}{\line(-1,0){0.12}}
\multiput(82.41,11.93)(-0.16,-0.12){32}{\line(-1,0){0.16}}
\multiput(77.28,8.10)(-0.21,-0.12){27}{\line(-1,0){0.21}}
\multiput(71.69,4.95)(-0.28,-0.11){21}{\line(-1,0){0.28}}
\multiput(65.76,2.55)(-0.44,-0.12){14}{\line(-1,0){0.44}}
\multiput(59.56,0.92)(-0.91,-0.12){7}{\line(-1,0){0.91}}
\put(53.20,0.10){\line(-1,0){6.41}}
\multiput(46.80,0.10)(-0.91,0.12){7}{\line(-1,0){0.91}}
\multiput(40.44,0.92)(-0.44,0.12){14}{\line(-1,0){0.44}}
\multiput(34.24,2.55)(-0.28,0.11){21}{\line(-1,0){0.28}}
\multiput(28.31,4.95)(-0.21,0.12){27}{\line(-1,0){0.21}}
\multiput(22.72,8.10)(-0.16,0.12){32}{\line(-1,0){0.16}}
\multiput(17.59,11.93)(-0.12,0.12){38}{\line(-1,0){0.12}}
\multiput(12.99,16.38)(-0.12,0.15){34}{\line(0,1){0.15}}
\multiput(8.99,21.39)(-0.12,0.20){28}{\line(0,1){0.20}}
\multiput(5.67,26.87)(-0.12,0.27){22}{\line(0,1){0.27}}
\multiput(3.08,32.73)(-0.11,0.38){16}{\line(0,1){0.38}}
\multiput(1.25,38.87)(-0.11,0.70){9}{\line(0,1){0.70}}
\multiput(0.23,45.20)(-0.10,3.20){2}{\line(0,1){3.20}}
\multiput(0.03,51.60)(0.10,1.06){6}{\line(0,1){1.06}}
\multiput(0.64,57.98)(0.12,0.52){12}{\line(0,1){0.52}}
\multiput(2.07,64.23)(0.12,0.32){19}{\line(0,1){0.32}}
\multiput(4.28,70.24)(0.12,0.23){25}{\line(0,1){0.23}}
\multiput(7.24,75.92)(0.12,0.17){31}{\line(0,1){0.17}}
\multiput(10.91,81.17)(0.12,0.13){36}{\line(0,1){0.13}}
\multiput(15.22,85.92)(0.14,0.12){35}{\line(1,0){0.14}}
\multiput(20.09,90.07)(0.18,0.12){30}{\line(1,0){0.18}}
\multiput(25.46,93.57)(0.24,0.12){24}{\line(1,0){0.24}}
\multiput(31.24,96.35)(0.36,0.12){17}{\line(1,0){0.36}}
\multiput(37.32,98.36)(0.91,0.12){14}{\line(1,0){0.91}}
\bezier{32}(31.00,106.66)(28.33,103.33)(32.33,104.00)
\bezier{20}(32.33,104.00)(35.00,104.33)(36.66,105.66)
\bezier{20}(36.66,105.66)(38.33,107.66)(36.33,108.33)
\put(5.33,2.33){}
\bezier{188}(1.33,61.33)(-4.67,52.67)(32.00,51.67)
\bezier{140}(32.00,51.67)(53.00,53.00)(65.00,60.67)
\bezier{168}(65.00,60.67)(92.67,77.33)(86.00,84.67)
\put(60.33,58.00){\vector(2,1){31.67}}
\put(31.67,111.67){\line(1,-3){37.22}}
\put(36.00,108.67){\vector(-4,1){1.33}}
\end{picture}
\end{center}
\caption{
\label{f-2004-f1}}
\end{figure}

In principle, such a setup could even measure the configuration of distant masses by Mach's principle.
Recall that, according to Einstein's perception of Mach,
the inertial motion of a body should be determined in relation to all
other bodies in the universe; in short, ``matter there governs inertia here.''
As the earth's gravity pull is known and the shift of falling bodies is measurable,
a reverse computation could yield the inertial motion the distant masses measurable by falling bodies.
But this is beyond the scope of this little review.

\section{Quantum mechanics}

Feyerabend wrote several contributions to the foundational debate
in quantum mechanics.
They are quite detailed and reflect the ongoing debate at the time they were written,
but I failed to find new aspects in them which had a lasting impact on the community.
At least Feyerabend was cautious enough not to state any erroneous claims as Popper.

Recall Feyerabend's statement cited above on people who have no idea why a stone falls to the ground
talking  with assurance about scientific method; where incompetence is accompanied by boredom
and self-righteousness.
These are very harsh,
critical words which in my opinion characterize Feyerabends (self-) provoking
stile.
They are, I think, not entirely unjust, and their the main premise in my opinion {\em is} correct:
most philosophers nowadays are at a complete loss
of understanding the more recent developments in physics.
With {\em philosophers} I mean everybody with an academic degree after
a study mainly concentrating on philosophy, as compared to the natural sciences.

There are great exceptions to the rule, but these are rare and sparse.
I certainly do not want to contribute to the ridiculous debate of the natural sciences with the rest of the faculties,
sparked by Sokal \cite{sokal-aff},
as I certainly do not want to argue that the natural sciences are immune to fraud,
misbehavior, stupidity and deception.
All I want to say is that any philosophy will be misleading
without a proper education in and knowledge of the  subject.
This is particularly true for the philosophy of science.
So, I am afraid, I have to urge philosophers and students of philosophy of science
to study mathematics, physics, logic, chemistry, biology and computer science proper.
At least the mastering of one of these subjects is necessary
in order to be able to comprehend, more so to contribute, to the ongoing debates in these areas.

In the meantime, physicists like myself will go wild and usurp territories
which would be better covered by
the philosophers, as they have much more background in the historical debates and are less inclined
to state ridiculously naive claims on foundational questions such as reality and metaphysics.
We desperately need philosophy after all, as we desperately need philosophers!
But we need to educate them better in the sciences, if they wish to consider science.
And please do not confuse attempts to brainwash people into science proper with concerns of competence.

From these very general remarks, let me now come back to quantum physics, which still remains
a very active research area.
Almost since its introduction in 1900 it has been the subject of intense philosophical debates,
both within the physics community --- the physicists, due to the good old Humboldt type curriculum,
were much better trained  in classical philosophy ---  and by laymen.
To this debate also Feyerabend contributed, as already noted.
If one is not willing to digest the volumes of Jammer
\cite{jammer:66,jammer1,jammer-92},
or the collection of original articles by Wheeler and Zurek \cite{wheeler-Zurek:83},
one gets a good glimpse of what was and still is going on from Schr\"odinger's
series of three articles on {\em ``Die gegenw{\"{a}}rtige {S}ituation in der {Q}uantenmechanik''} \cite{schrodinger}
(English translation {\em ``The Present Situation In Quantum Mechanics''} \cite[pp. 152-167]{wheeler-Zurek:83}).
I think that I can safely say that, although {\em ``nobody understands quantum mechanics''}
(cf. Richard Feynman in Ref.~\cite{feynman-law}, p. 129),
nobody not able to comprehend these Schr\"odinger articles should make a public appearance
on related topics.

\subsection{Old topics in a new terminology: scholasticism, realism--idealism}

The debate on the foundations of quantum mechanics has taken a somewhat unexpected twist,
in particular in recent years, when due to new experimental techniques single quantum events
could be investigated:
it turned into controversies with associated long-lasting debates and  huge philosophical records.
At the same time, some of the old concepts became formalized.
In what follows, I take up the task of reviewing some of these issues,
being well aware of the risk of being dilettantish.

One of the big issues is related to the realism {\it versus} idealism debate.
Stace \cite{stace} characterized realism by the
supposition that {\em ``some entities sometimes exist without being experienced by any finite mind.''}
He was an outspoken idealist, claiming that
{\em ``$\ldots$
we have not the faintest reason for believing in the existence of
inexperienced entities
$\ldots$
[[Realism]] has been adopted
$\ldots$
solely because it simplifies our view of the universe.''}

In quantum mechanics, this debate may probably be summed up by the term
{\em Bohr--Einstein debate,}
with Einstein firmly positioned as a realist.
One of the questions concerns the existence of physical properties
even in the absence of their direct physical observation.
Einstein \cite{epr}
suggested to do just that, and effectively accept indirectly inferred counterfactuals
als {\em ``elements of physical reality.''}

A further step was taken by the famous Swizz mathematician Specker,
who, stimulated by the quantum logic developed by Birkhoff and von Neumann
\cite{birkhoff-36},
pondered about the logic of propositions which are not co-measurable; i.e., not simultaneously
measurable \cite{specker-60}.
Specker related such structures in quantum physics to
  Scholasticism,
in particular to scholastic speculations  about the existence of ``infuturabilities'' or
``counterfactuals.''
The question, for instance posed by Saint Thomas Aquinas, is whether or not
the omniscience (comprehensive knowledge) of God extends to events which
would have occurred if something  had happened which did not
happen.
If so, could all such events be pasted together to form a consistent whole?

Concerns about co-measurability were inevitable because  quantum mechanics has introduced new features
hitherto hardly heard of in classical physics.
Complementarity was a consequence of Heisenberg's formulation of the theory.
Non-commutative operators and the resulting non-distributive propositional structure
became facts of everyday professional life for theoreticians and experimentalists alike.
Yet, what is hardly noticed even by the specialists is the fact
that complementarity and non-distributivity
not necessarily implies total abandonment of non-classicality:
quasi-classical models such as generalized urn models
\cite{wright:pent,wright} and finite automata (e.g., Chapter 10 of Ref.~\cite{svozil-ql})
can be isomorphically embedded into Boolean algebras
with the help of two-valued probability measures,
which abound for such models  \cite{svozil-2001-eua}.

That such non-co-measurable propositions exist might have come as a surprise
even to the classical mind in retrospect.
Indeed, this is a good example for the fact that our phantasy is not good and wild enough to conceive of
the many available alternative options we have for almost any given situation.

Bell \cite{bell-87} and others took up an idea expressed
in an article of the late Einstein, co-published with Podolsky and Rosen \cite{epr}.
The Bell-type inequalities are particular instances of Boole's
{\em ``conditions of physical existence''} \cite{Boole,Boole-62},
which are consistency conditions on joint probabilities,
for specific physical setups.
The argument  Einstein-Podolsky-Rosen argument \cite{epr}
suggests that, although two non-co-measurable properties
(associated with complementary observables)
cannot be directly measured at a single quantum,
one may infer from certain two-quanta states
(satisfying a uniqueness property \cite{svozil-2004-vax})
one property per quantum and subsequently counterfactually
infer the other property from its twin quantum.

In doing so one implicitly assumes that counterfactuals exist:
it is assumed that, if the counterfactually inferred property would have been measured
---
which was not the case
---
it would have come out in the expected way.

On top of that,
the Bell inequalities contain sums of terms (e.g., joint probabilities or expectation values)
which
could only be measured subsequently (or at least in different experimental setups; one at a time),
since they correspond to different parameter settings which, according to the quantum complementarity rules,
especially complementarity, cannot be measured simultaneously.
To the realist assuming that
entities exist without being experienced by any finite mind,
this is no big deal.
Even collecting terms associated with measuring different non-co-measurable setups
and summing them up as if they referred to a single quantum is hardly disturbing.
(This makes possible a criticism put forward recently \cite{Hess&Philipp2002}.)

But the assumption of an ``all out'' omni-realism may be improper in the quantum domain.
Quanta prepared in a specific state in a given experimental context
might simply not be capable to ``know''
their precise states in different contexts \cite{svozil-2003-garda}.

As an analogy, no finite agent such as a computer program can be set up to answer
all conceivable questions  --- it may be at a complete loss at answering some or even most of them.
This kind of restricted way appears quite natural and is not very exciting; certainly not
as ``mindboggling'' or ``mystical'' as the quantum tales of Bohr.
It is just another kind of realism,
one which is based on the assumption that certain things do not have all conceivable properties
we would wish them to have; just a finite number of properties, that is it.

Nevertheless, let me emphasize that non-classicality of quantum mechanics goes well beyond complementarity.
There is a finite constructive proof of the impossibility
of value definiteness for quantized systems whose description require Hilbert spaces
of dimension higher than two.
It turned out that formally there are
``not enough'' two-valued states to allow a faithful embedding of certain tightly interconnected
finite propositional structures
into any Boolean algebra.
This can be characterized by
non-separable or non-unital sets of two-valued states;
the strongest result being the nonexistence of two-valued states today known as the ``Kochen-Specker theorem''
\cite{kochen1}.
In fact, once the propositional structure has been enumerated explicitly, a proof is technically not very demanding
and amounts to a coloring theorem (chromaticity) on these sets.

\begin{figure}
\begin{center}
\unitlength 0.50mm
\linethickness{0.4pt}
\begin{picture}(190.67,109.67)
\multiput(165.67,19.67)(-0.12,0.12){167}{\line(0,1){0.12}}
\put(145.67,39.67){\line(0,1){40.00}}
\multiput(145.67,79.67)(0.12,0.12){167}{\line(0,1){0.12}}
\multiput(165.67,99.67)(0.12,-0.12){167}{\line(1,0){0.12}}
\put(185.67,79.67){\line(0,-1){40.00}}
\multiput(185.67,39.67)(-0.12,-0.12){167}{\line(-1,0){0.12}}
\put(185.34,59.67){\line(-1,0){39.67}}
\put(185.67,39.67){\circle{2.00}}
\put(185.67,59.67){\circle{2.00}}
\put(185.67,79.67){\circle{2.00}}
\put(145.67,39.67){\circle{2.00}}
\put(145.67,59.67){\circle{2.00}}
\put(145.67,79.67){\circle{2.00}}
\put(165.67,19.67){\circle{2.00}}
\put(165.67,99.67){\circle{2.00}}
\multiput(165.67,19.67)(-0.21,0.12){334}{\line(-1,0){0.21}}
\multiput(95.67,59.67)(0.21,0.12){334}{\line(1,0){0.21}}
\put(95.67,59.67){\circle{2.00}}
\put(95.67,74.67){\makebox(0,0)[cc]{$a_8=a_8'$}}
\put(165.67,109.67){\makebox(0,0)[cc]{$b=a_9=a_0'$}}
\put(140.67,79.67){\makebox(0,0)[cc]{$a_2'$}}
\put(140.67,59.67){\makebox(0,0)[cc]{$a_6'$}}
\put(140.67,39.67){\makebox(0,0)[cc]{$a_4'$}}
\put(190.67,39.67){\makebox(0,0)[cc]{$a_3'$}}
\put(190.67,59.67){\makebox(0,0)[cc]{$a_5'$}}
\put(190.67,80.00){\makebox(0,0)[cc]{$a_1'$}}
\put(165.67,9.67){\makebox(0,0)[cc]{$a_7'$}}
\multiput(25.00,19.67)(0.12,0.12){167}{\line(0,1){0.12}}
\put(45.00,39.67){\line(0,1){40.00}}
\multiput(45.00,79.67)(-0.12,0.12){167}{\line(0,1){0.12}}
\multiput(25.00,99.67)(-0.12,-0.12){167}{\line(-1,0){0.12}}
\put(5.00,79.67){\line(0,-1){40.00}}
\multiput(5.00,39.67)(0.12,-0.12){167}{\line(1,0){0.12}}
\put(5.33,59.67){\line(1,0){39.67}}
\put(5.00,39.67){\circle{2.00}}
\put(5.00,59.67){\circle{2.00}}
\put(5.00,79.67){\circle{2.00}}
\put(45.00,39.67){\circle{2.00}}
\put(45.00,59.67){\circle{2.00}}
\put(45.00,79.67){\circle{2.00}}
\put(25.00,19.67){\circle{2.00}}
\put(25.00,99.67){\circle{2.00}}
\multiput(25.00,19.67)(0.21,0.12){334}{\line(1,0){0.21}}
\multiput(95.00,59.67)(-0.21,0.12){334}{\line(-1,0){0.21}}
\put(25.00,109.67){\makebox(0,0)[cc]{$a=a_0=a_9'$}}
\put(50.00,79.67){\makebox(0,0)[cc]{$a_2$}}
\put(50.00,59.67){\makebox(0,0)[cc]{$a_6$}}
\put(50.00,39.67){\makebox(0,0)[cc]{$a_4$}}
\put(-0.00,39.67){\makebox(0,0)[cc]{$a_3$}}
\put(-0.00,59.67){\makebox(0,0)[cc]{$a_5$}}
\put(-0.00,80.00){\makebox(0,0)[cc]{$a_1$}}
\put(25.00,9.67){\makebox(0,0)[cc]{$a_7$}}
\end{picture}
\end{center}
\caption{
Greechie (orthogonality) diagram
\cite{greechie:71},  consisting of {\em points} which
symbolize observables (representable by the spans of vectors
in $n$-dimensional Hilbert space).
Any $n$ points belonging to a context; i.e., to a maximal set of co-measurable observables
(representable as some orthonormal basis of  $n$-dimensional Hilbert space),
are connected by {\em smooth curves}.
Two smooth curves may be crossing in  common {\em link observables}.
In three dimensions, smooth curves and the associated points stand for tripods.
f $P(a=a_0=a_9') = 1$ for any
two-valued probability measure $P$, then $P(a_8) =0$.
Furthermore,
$P(a_7)=0$, since by a similar argument $P(a)=1$ implies $P(a_7)=0$.
Therefore, $P(b=a_9=a_0') = 1$. Symmetry requires that the
reverse implication is also fulfilled, and therefore  $P(b) =
P(a)$ for every two-valued probability measure $P$.
\label{f-2004-fey2}
}
\end{figure}
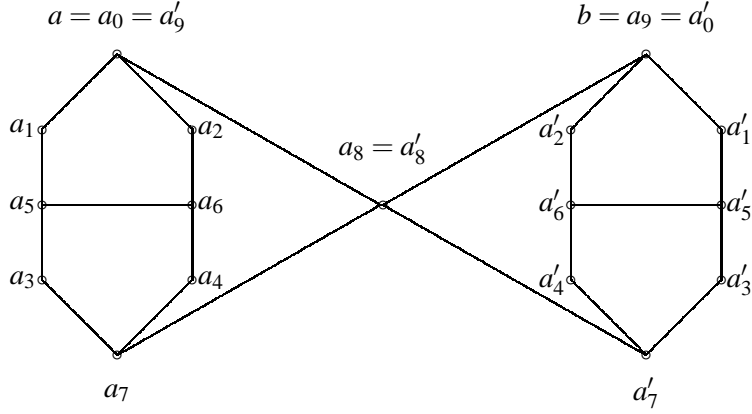

To get a taste of the type of argument, Fig.~\ref{f-2004-fey2} depicts a quantum propositional structure; i.e., a logic,
with a  non-separable set of two-valued states, such that $P(a) = P(b)$.
It is the graph $\Gamma_3$ of Ref.~\cite{kochen1}
represented by the Greechie orthogonality diagram in Ref.~\cite{svozil-tkadlec}.

\subsection{Physicists at their own}

So what have the physicists produced when they
were left alone to interpret the ``meaning'' of the formalism?
They have developed a variety of interpretations,
the number of which is probably as great as there are physicists
and almost as great as the number of philosophical concepts of reality.
Let us just mention a couple of these interpretations and some of their creators and devotees:
{\em
the Copenhagen interpretation (Bohr),
the many-worlds interpretation  (Everett),
Bohm's interpretation,
the consistent histories approach (Griffiths) and finally a
``Realistic'' interpretation  (Einstein, De Broglie, Schr\"odinger).}
All of these offer no difference in the predictions and formalization of
quantum mechanics.
Yet they serve as a kind of scaffolding \cite{svozil-2002-noiq};
without it science would be reduced to an automated proof technique,
straying without guidance,
devoid of any
idea of how to proceed with (hopefully) progressive research programs
\cite{lakatosch}.

Let me also mention a somewhat unrelated issue.
Some physicists ``go wild'' and pretend that the transient status of their science reflects
final truth of the world; they
tell fairy tales about the first three minutes of the Universe, short histories of time and what not.
This is good for marketing purposes and sells well.
What they do not seem to acknowledge and the public simply does not want to hear
is the historic aspect of our findings which makes our present knowledge transient
and preliminary.

In this way of thinking, which emphasizes transitions and the continuation of research programs,
I tend to agree with Lakatosch \cite{lakatosch}.
Feyerabend's critique of Lakatosch is that the latter does not offer a methodology.
Yet the same could be said of Feyerabend's methodology
\cite{feyerabend-defense,feyer-81}.
And if openness, or suspended attention as Freud put it, is no methodology, then so be it.

\section{Some personal remarks}

In the spring semester 1983, I  attended a course of Feyerabend on the philosophy
of science in Berkeley during my stay as a visiting scholar at the Lawrence Berkeley
Laboratory and the University.
Feyerabend made a sad but rebellious impression,
best described by the German word ``unerf\"ullt,'' whose English translation
is ``unrealised, unfulfilled.''
His spirits were strong and he gave a quite good performance.

His audience consisted of about twenty people; maybe half of them students,
the other part devotees and curious listeners.
It was rather obvious from his reactions that he despised the fan club gathered
to listen to the master's voice, but somehow he longed for them, too---very ambivalent,
but probably not unusual for prominent people.

After one of his lectures, I approached him and asked him if we could meet.
He responded friendly but not very enthusiastically.
I guess he was not really interested in a very young, naive physicist from Vienna.
What cold I offer him despite boredom?

Alas, I had the impression that he was after the girls.
I guess that if I would have been a pretty girl,
I would have had very good chances of meeting him and have a chat or two.
But again this is one of those counterfactuals I was speaking of before.

Hardly anybody despite Professor Fischer \cite{fischer} and Feyerabend himself
\cite{feyerabend-auto}
spoke about Feyerabend's sex life.
For me it is quite remarkable that he was a womanizer on one side
while on the other hand seemed not to have had a single coitus during his entire life span.
I take this as an indication that there was deep dissatisfaction with this situation,
a malady which was possibly not only caused by his war time injuries,
but by psychic traumata which may have been deeply hidden and never showed up during
his conscious phases in-between dreams and deep sleep.
This may also be the ultimate reason for the kind of ``Unerf\"ulltheit''
I observed, but of course I am wildly speculating here.

There was another problem with his thinking which Feyerabend took quite light-heartedly,
at least that was my impression:
nobody took him seriously.
Paul Feyerabend had become almost a shooting star of philosophy of science,
an icon of freedom and heresy
to a generation coming of age in the late period of the twentieth century.
Yet he never quite managed to obtain influence
and convince scientific peers, governments and electorates to implement
his recommendations regarding the selection of science funding and the implementation
of science in general.
In the lectures I attended during 1983,
he strongly supported a system of  lay judges for science assessment  and  financing;
in closest analogy to the procedures established in the judicial system.
I did never find so strong commitments to  lay judges in his writings
as I heard them in these lectures.

There may be very simple reasons for not taking Feyerabend too seriously:
Feyerabend's proposal would introduce an uncontrollable element in the distribution of money.
For not only might quacks receive public funding;
An even more disturbing consequence could be that, by a fairly independent
selection of committee members and lay evaluators,
powerful groups within the scientific community might loose their carefully
crafted and delicately executed influence over the smooth flow of money towards them and their clients.

To some readers this may sound like a silly conspiracy theory.
To this I respond that the matter is not obvious but
quite serious and deserves careful attention of the general tax paying
public, to which this article is not addressed.
Let me just mention
a large-scale study \cite{1981-cole}, in which 150 research projects of
physics, chemistry and economic science were re-examined by the {\it
National Science Foundation.} The results were devastating.
This study showed how strongly the acceptance or refusal of
a research project depends on the choice of the particular reviewer
evaluating that proposal:  {\em ``An experiment in which 150
proposals submitted to the National Science Foundation were evaluated
independently by a new set of reviewers indicates that getting a
research grant depends to a significant extend on chance.
$\ldots$
the degree of disagreement within the
population of eligible reviewers is such that whether or not a proposal
is funded depends in a large proportion of cases upon which reviewers
happen to be selected for it.''}
This finding is rather disturbing, as
in many funding agencies, the ``referents'' in charge
of selection of the peer reviewers
are nominated in a rather unaccountable and
certainly not very transparent  manner.

There is one development which Feyerabend did not foresee: the growing detrimental dominance
of the administrative bureaucracies over the scientists.
This is administered through an ever increasing net of what appears to be checks and balances,
of scientometric factors and numbers, of frequent evaluations and proposals
and various certification and standardization procedures.
This makes perfect sense for the administration,
whose major task it is to distribute public money smoothly, accountably in terms of records,  and
free of risks.
But this is not seldom opposed to science, and also opposed to the methodological
openness Feyerabend had in mind.
It also is quite frustrating for the researchers which are captives of this treadmill.
The behemoth created by the Sixth Framework Programme (FP6)
and the establishment of the European Research Area (ERA)
are such examples, but there are numerous others on local and institutional scales.

Let me finish more conciliatory and stress the heritage of Paul Feyerabend.
To me, of the many wise and weird things he said and wrote, two messages
are very important.
The first one is in the spirit of the
Enlightenment and gets close to what also Kant had in mind:
try on your own, let not others decide what you think;
do not stop where other people, authorities  and  mandarins tell you to halt.
And finally, in his last manuscript, Feyerabend calls upon us
to reach out and conquer the abundance.


\begin{thebibliography}{42}
\expandafter\ifx\csname natexlab\endcsname\relax\def\natexlab#1{#1}\fi
\expandafter\ifx\csname bibnamefont\endcsname\relax
  \def\bibnamefont#1{#1}\fi
\expandafter\ifx\csname bibfnamefont\endcsname\relax
  \def\bibfnamefont#1{#1}\fi
\expandafter\ifx\csname citenamefont\endcsname\relax
  \def\citenamefont#1{#1}\fi
\expandafter\ifx\csname url\endcsname\relax
  \def\url#1{\texttt{#1}}\fi
\expandafter\ifx\csname urlprefix\endcsname\relax\def\urlprefix{URL }\fi
\providecommand{\bibinfo}[2]{#2}
\providecommand{\eprint}[2][]{\url{#2}}

\bibitem[{\citenamefont{Feyerabend}(1995)}]{feyerabend-auto}
\bibinfo{author}{\bibfnamefont{P.~K.} \bibnamefont{Feyerabend}},
  \emph{\bibinfo{title}{Killing time}} (\bibinfo{publisher}{The University of
  Chicago Press}, \bibinfo{address}{Chicago and London}, \bibinfo{year}{1995}).

\bibitem[{\citenamefont{Peres}(2002)}]{2002-peres}
\bibinfo{author}{\bibfnamefont{A.}~\bibnamefont{Peres}},
  \bibinfo{journal}{Stud. History and Philos. of Modern Physics}
  \textbf{\bibinfo{volume}{33}}, \bibinfo{pages}{23} (\bibinfo{year}{2002}),
  \urlprefix\url{http://dx.doi.org/10.1016/S1355-2198(01)00034-X}.

\bibitem[{\citenamefont{Chiara}(2002)}]{dalla-2002}
\bibinfo{author}{\bibfnamefont{M.~D.} \bibnamefont{Chiara}}
  (\bibinfo{year}{2002}), \bibinfo{note}{invited talk at the Karl Popper 2002
  Centenary Congress, Vienna, July 5th}.

\bibitem[{\citenamefont{Fischer}(1999)}]{fischer}
\bibinfo{author}{\bibfnamefont{K.~R.} \bibnamefont{Fischer}}, in
  \emph{\bibinfo{booktitle}{Aufs{\"{a}}tze zur angloamerikanischen und
  {\"{o}}sterreichischen Philosophie}}, edited by
  \bibinfo{editor}{\bibfnamefont{K.~R.} \bibnamefont{Fischer}}
  (\bibinfo{publisher}{Lang}, \bibinfo{address}{Frankfurt am Main, Wien},
  \bibinfo{year}{1999}), pp. \bibinfo{pages}{77--89}.

\bibitem[{\citenamefont{Theocharis and Psimoloulos}(1987)}]{theo-psi-1987}
\bibinfo{author}{\bibfnamefont{T.}~\bibnamefont{Theocharis}} \bibnamefont{and}
  \bibinfo{author}{\bibfnamefont{M.}~\bibnamefont{Psimoloulos}},
  \bibinfo{journal}{Nature} \textbf{\bibinfo{volume}{329}},
  \bibinfo{pages}{595} (\bibinfo{year}{1987}).

\bibitem[{\citenamefont{Svozil}(2002{\natexlab{a}})}]{svozil-2002-popper}
\bibinfo{author}{\bibfnamefont{K.}~\bibnamefont{Svozil}}
  (\bibinfo{year}{2002}{\natexlab{a}}), \eprint{physics/0207115}.

\bibitem[{\citenamefont{Feyerabend}(1975)}]{feyerabend-defense}
\bibinfo{author}{\bibfnamefont{P.~K.} \bibnamefont{Feyerabend}},
  \bibinfo{journal}{Radical Philosophy} \textbf{\bibinfo{volume}{11}}
  (\bibinfo{year}{1975}), \bibinfo{note}{reprinted in \cite{feyer-81}}.

\bibitem[{\citenamefont{Feyerabend}(1981)}]{feyer-81}
\bibinfo{author}{\bibfnamefont{P.~K.} \bibnamefont{Feyerabend}}, in
  \emph{\bibinfo{booktitle}{Scientific Revolutions}}, edited by
  \bibinfo{editor}{\bibfnamefont{I.}~\bibnamefont{Hacking}}
  (\bibinfo{publisher}{Oxford University Press}, \bibinfo{address}{Oxford},
  \bibinfo{year}{1981}), pp. \bibinfo{pages}{156--167}.

\bibitem[{\citenamefont{Feyerabend}(1974)}]{feyerabend}
\bibinfo{author}{\bibfnamefont{P.~K.} \bibnamefont{Feyerabend}},
  \emph{\bibinfo{title}{Against Method}} (\bibinfo{publisher}{New Left Books},
  \bibinfo{address}{London}, \bibinfo{year}{1974}).

\bibitem[{\citenamefont{Hall}(1903{\natexlab{a}})}]{hall-1903a}
\bibinfo{author}{\bibfnamefont{E.~H.} \bibnamefont{Hall}},
  \bibinfo{journal}{Phys. Rev. (Series I)} \textbf{\bibinfo{volume}{17}},
  \bibinfo{pages}{179} (\bibinfo{year}{1903}{\natexlab{a}}),
  \urlprefix\url{http://dx.doi.org/10.1103/PhysRevSeriesI.17.179}.

\bibitem[{\citenamefont{Hall}(1903{\natexlab{b}})}]{hall-1903b}
\bibinfo{author}{\bibfnamefont{E.~H.} \bibnamefont{Hall}},
  \bibinfo{journal}{Phys. Rev. (Series I)} \textbf{\bibinfo{volume}{17}},
  \bibinfo{pages}{245} (\bibinfo{year}{1903}{\natexlab{b}}),
  \urlprefix\url{http://dx.doi.org/10.1103/PhysRevSeriesI.17.245}.

\bibitem[{\citenamefont{Armitage}(1941-47)}]{armitage}
\bibinfo{author}{\bibfnamefont{A.}~\bibnamefont{Armitage}},
  \bibinfo{journal}{Annals of Science} \textbf{\bibinfo{volume}{5}},
  \bibinfo{pages}{342} (\bibinfo{year}{1941-47}).

\bibitem[{\citenamefont{Sokal}()}]{sokal-aff}
\bibinfo{author}{\bibfnamefont{A.}~\bibnamefont{Sokal}},
  \bibinfo{note}{http://www.physics.nyu.edu/faculty/sokal/},
  \urlprefix\url{http://www.physics.nyu.edu/faculty/sokal/}.

\bibitem[{\citenamefont{Jammer}(1966)}]{jammer:66}
\bibinfo{author}{\bibfnamefont{M.}~\bibnamefont{Jammer}},
  \emph{\bibinfo{title}{The Conceptual Development of Quantum Mechanics}}
  (\bibinfo{publisher}{McGraw-Hill Book Company}, \bibinfo{address}{New York},
  \bibinfo{year}{1966}).

\bibitem[{\citenamefont{Jammer}(1974)}]{jammer1}
\bibinfo{author}{\bibfnamefont{M.}~\bibnamefont{Jammer}},
  \emph{\bibinfo{title}{The Philosophy of Quantum Mechanics}}
  (\bibinfo{publisher}{John Wiley \& Sons}, \bibinfo{address}{New York},
  \bibinfo{year}{1974}).

\bibitem[{\citenamefont{Jammer}(1992)}]{jammer-92}
\bibinfo{author}{\bibfnamefont{M.}~\bibnamefont{Jammer}}, in
  \emph{\bibinfo{booktitle}{Bell's Theorem and the Foundations of Modern
  Physics}}, edited by \bibinfo{editor}{\bibfnamefont{A.}~\bibnamefont{van~der
  Merwe}}, \bibinfo{editor}{\bibfnamefont{F.}~\bibnamefont{Selleri}},
  \bibnamefont{and} \bibinfo{editor}{\bibfnamefont{G.}~\bibnamefont{Tarozzi}}
  (\bibinfo{publisher}{World Scientific}, \bibinfo{address}{Singapore},
  \bibinfo{year}{1992}), pp. \bibinfo{pages}{1--23}.

\bibitem[{\citenamefont{Wheeler and Zurek}(1983)}]{wheeler-Zurek:83}
\bibinfo{author}{\bibfnamefont{J.~A.} \bibnamefont{Wheeler}} \bibnamefont{and}
  \bibinfo{author}{\bibfnamefont{W.~H.} \bibnamefont{Zurek}},
  \emph{\bibinfo{title}{Quantum Theory and Measurement}}
  (\bibinfo{publisher}{Princeton University Press},
  \bibinfo{address}{Princeton}, \bibinfo{year}{1983}).

\bibitem[{\citenamefont{Schr{\"{o}}dinger}(1935)}]{schrodinger}
\bibinfo{author}{\bibfnamefont{E.}~\bibnamefont{Schr{\"{o}}dinger}},
  \bibinfo{journal}{Naturwissenschaften} \textbf{\bibinfo{volume}{23}},
  \bibinfo{pages}{807} (\bibinfo{year}{1935}), \bibinfo{note}{{E}nglish
  translation in \cite{trimmer} and \cite[pp. 152-167]{wheeler-Zurek:83};
  http://www.emr.hibu.no/lars/eng/cat/},
  \urlprefix\url{http://www.emr.hibu.no/lars/eng/cat/}.

\bibitem[{\citenamefont{Feynman}(1965)}]{feynman-law}
\bibinfo{author}{\bibfnamefont{R.~P.} \bibnamefont{Feynman}},
  \emph{\bibinfo{title}{The Character of Physical Law}}
  (\bibinfo{publisher}{MIT Press}, \bibinfo{address}{Cambridge, MA},
  \bibinfo{year}{1965}).

\bibitem[{\citenamefont{Stace}(1949)}]{stace}
\bibinfo{author}{\bibfnamefont{W.~T.} \bibnamefont{Stace}}, in
  \emph{\bibinfo{booktitle}{Readings in philosophical analysis}}, edited by
  \bibinfo{editor}{\bibfnamefont{H.}~\bibnamefont{Feigl}} \bibnamefont{and}
  \bibinfo{editor}{\bibfnamefont{W.}~\bibnamefont{Sellars}}
  (\bibinfo{publisher}{Appleton--Century--Crofts}, \bibinfo{address}{New York},
  \bibinfo{year}{1949}).

\bibitem[{\citenamefont{Einstein et~al.}(1935)\citenamefont{Einstein, Podolsky,
  and Rosen}}]{epr}
\bibinfo{author}{\bibfnamefont{A.}~\bibnamefont{Einstein}},
  \bibinfo{author}{\bibfnamefont{B.}~\bibnamefont{Podolsky}}, \bibnamefont{and}
  \bibinfo{author}{\bibfnamefont{N.}~\bibnamefont{Rosen}},
  \bibinfo{journal}{Physical Review} \textbf{\bibinfo{volume}{47}},
  \bibinfo{pages}{777} (\bibinfo{year}{1935}),
  \urlprefix\url{http://dx.doi.org/10.1103/PhysRev.47.777}.

\bibitem[{\citenamefont{Birkhoff and von Neumann}(1936)}]{birkhoff-36}
\bibinfo{author}{\bibfnamefont{G.}~\bibnamefont{Birkhoff}} \bibnamefont{and}
  \bibinfo{author}{\bibfnamefont{J.}~\bibnamefont{von Neumann}},
  \bibinfo{journal}{Annals of Mathematics} \textbf{\bibinfo{volume}{37}},
  \bibinfo{pages}{823} (\bibinfo{year}{1936}).

\bibitem[{\citenamefont{Specker}(1960)}]{specker-60}
\bibinfo{author}{\bibfnamefont{E.}~\bibnamefont{Specker}},
  \bibinfo{journal}{Dialectica} \textbf{\bibinfo{volume}{14}},
  \bibinfo{pages}{175} (\bibinfo{year}{1960}), \bibinfo{note}{reprinted in
  \cite[pp. 175--182]{specker-ges}; {E}nglish translation: {\it The logic of
  propositions which are not simultaneously decidable}, reprinted in \cite[pp.
  135-140]{hooker}}.

\bibitem[{\citenamefont{Wright}(1978)}]{wright:pent}
\bibinfo{author}{\bibfnamefont{R.}~\bibnamefont{Wright}}, in
  \emph{\bibinfo{booktitle}{Mathematical Foundations of Quantum Theory}},
  edited by \bibinfo{editor}{\bibfnamefont{A.~R.} \bibnamefont{Marlow}}
  (\bibinfo{publisher}{Academic Press}, \bibinfo{address}{New York},
  \bibinfo{year}{1978}), pp. \bibinfo{pages}{255--274}.

\bibitem[{\citenamefont{Wright}(1990)}]{wright}
\bibinfo{author}{\bibfnamefont{R.}~\bibnamefont{Wright}},
  \bibinfo{journal}{Foundations of Physics} \textbf{\bibinfo{volume}{20}},
  \bibinfo{pages}{881} (\bibinfo{year}{1990}).

\bibitem[{\citenamefont{Svozil}(1998)}]{svozil-ql}
\bibinfo{author}{\bibfnamefont{K.}~\bibnamefont{Svozil}},
  \emph{\bibinfo{title}{Quantum Logic}} (\bibinfo{publisher}{Springer},
  \bibinfo{address}{Singapore}, \bibinfo{year}{1998}).

\bibitem[{\citenamefont{Svozil}(2002{\natexlab{b}})}]{svozil-2001-eua}
\bibinfo{author}{\bibfnamefont{K.}~\bibnamefont{Svozil}}
  (\bibinfo{year}{2002}{\natexlab{b}}), \eprint{quant-ph/0209136}.

\bibitem[{\citenamefont{Bell}(1987)}]{bell-87}
\bibinfo{author}{\bibfnamefont{J.~S.} \bibnamefont{Bell}},
  \emph{\bibinfo{title}{Speakable and Unspeakable in Quantum Mechanics}}
  (\bibinfo{publisher}{Cambridge University Press},
  \bibinfo{address}{Cambridge}, \bibinfo{year}{1987}).

\bibitem[{\citenamefont{Boole}(1958)}]{Boole}
\bibinfo{author}{\bibfnamefont{G.}~\bibnamefont{Boole}},
  \emph{\bibinfo{title}{An investigation of the laws of thought}}
  (\bibinfo{publisher}{Dover edition}, \bibinfo{address}{New York},
  \bibinfo{year}{1958}).

\bibitem[{\citenamefont{Boole}(1862)}]{Boole-62}
\bibinfo{author}{\bibfnamefont{G.}~\bibnamefont{Boole}},
  \bibinfo{journal}{Philosophical Transactions of the Royal Society of London}
  \textbf{\bibinfo{volume}{152}}, \bibinfo{pages}{225} (\bibinfo{year}{1862}).

\bibitem[{\citenamefont{Svozil}(2004{\natexlab{a}})}]{svozil-2004-vax}
\bibinfo{author}{\bibfnamefont{K.}~\bibnamefont{Svozil}}
  (\bibinfo{year}{2004}{\natexlab{a}}), \eprint{quant-ph/0406014}.

\bibitem[{\citenamefont{Hess and Philipp}(2002)}]{Hess&Philipp2002}
\bibinfo{author}{\bibfnamefont{K.}~\bibnamefont{Hess}} \bibnamefont{and}
  \bibinfo{author}{\bibfnamefont{W.}~\bibnamefont{Philipp}},
  \bibinfo{journal}{Europhysics Letters} \textbf{\bibinfo{volume}{57}},
  \bibinfo{pages}{775} (\bibinfo{year}{2002}).

\bibitem[{\citenamefont{Svozil}(2004{\natexlab{b}})}]{svozil-2003-garda}
\bibinfo{author}{\bibfnamefont{K.}~\bibnamefont{Svozil}},
  \bibinfo{journal}{Journal of Modern Optics} \textbf{\bibinfo{volume}{51}},
  \bibinfo{pages}{811} (\bibinfo{year}{2004}{\natexlab{b}}),
  \eprint{quant-ph/0308110}.

\bibitem[{\citenamefont{Kochen and Specker}(1967)}]{kochen1}
\bibinfo{author}{\bibfnamefont{S.}~\bibnamefont{Kochen}} \bibnamefont{and}
  \bibinfo{author}{\bibfnamefont{E.~P.} \bibnamefont{Specker}},
  \bibinfo{journal}{Journal of Mathematics and Mechanics}
  \textbf{\bibinfo{volume}{17}}, \bibinfo{pages}{59} (\bibinfo{year}{1967}),
  \bibinfo{note}{reprinted in \cite[pp. 235--263]{specker-ges}}.

\bibitem[{\citenamefont{Svozil and Tkadlec}(1996)}]{svozil-tkadlec}
\bibinfo{author}{\bibfnamefont{K.}~\bibnamefont{Svozil}} \bibnamefont{and}
  \bibinfo{author}{\bibfnamefont{J.}~\bibnamefont{Tkadlec}},
  \bibinfo{journal}{Journal of Mathematical Physics}
  \textbf{\bibinfo{volume}{37}}, \bibinfo{pages}{5380} (\bibinfo{year}{1996}),
  \urlprefix\url{http://dx.doi.org/10.1063/1.531710}.

\bibitem[{\citenamefont{Greechie}(1971)}]{greechie:71}
\bibinfo{author}{\bibfnamefont{J.~R.} \bibnamefont{Greechie}},
  \bibinfo{journal}{Journal of Combinatorial Theory}
  \textbf{\bibinfo{volume}{10}}, \bibinfo{pages}{119} (\bibinfo{year}{1971}).

\bibitem[{\citenamefont{Svozil}(2002{\natexlab{c}})}]{svozil-2002-noiq}
\bibinfo{author}{\bibfnamefont{K.}~\bibnamefont{Svozil}}
  (\bibinfo{year}{2002}{\natexlab{c}}), \eprint{quant-ph/0204168}.

\bibitem[{\citenamefont{Lakatos}(1978)}]{lakatosch}
\bibinfo{author}{\bibfnamefont{I.}~\bibnamefont{Lakatos}},
  \emph{\bibinfo{title}{Philosophical Papers. 1.~The Methodology of Scientific
  Research Programmes}} (\bibinfo{publisher}{Cambridge University Press},
  \bibinfo{address}{Cambridge}, \bibinfo{year}{1978}).

\bibitem[{\citenamefont{Cole et~al.}(1981)\citenamefont{Cole, Cole, and
  Simon}}]{1981-cole}
\bibinfo{author}{\bibfnamefont{S.}~\bibnamefont{Cole}},
  \bibinfo{author}{\bibfnamefont{J.~R.} \bibnamefont{Cole}}, \bibnamefont{and}
  \bibinfo{author}{\bibfnamefont{G.~A.} \bibnamefont{Simon}},
  \bibinfo{journal}{Science} \textbf{\bibinfo{volume}{214}},
  \bibinfo{pages}{881} (\bibinfo{year}{1981}).

\bibitem[{\citenamefont{Trimmer}(1980)}]{trimmer}
\bibinfo{author}{\bibfnamefont{J.~D.} \bibnamefont{Trimmer}},
  \bibinfo{journal}{Proc. Am. Phil. Soc.} \textbf{\bibinfo{volume}{124}},
  \bibinfo{pages}{323} (\bibinfo{year}{1980}), \bibinfo{note}{reprinted in
  \cite[pp. 152-167]{wheeler-Zurek:83}.}

\bibitem[{\citenamefont{Specker}(1990)}]{specker-ges}
\bibinfo{author}{\bibfnamefont{E.}~\bibnamefont{Specker}},
  \emph{\bibinfo{title}{Selecta}} (\bibinfo{publisher}{Birkh{\"{a}}user
  Verlag}, \bibinfo{address}{Basel}, \bibinfo{year}{1990}).

\bibitem[{\citenamefont{Hooker}(1975)}]{hooker}
\bibinfo{author}{\bibfnamefont{C.~A.} \bibnamefont{Hooker}}, in
  \emph{\bibinfo{booktitle}{The Logico-Algebraic Approach to Quantum Mechanics.
  Volume I: Historical Evolution}} (\bibinfo{publisher}{Reidel},
  \bibinfo{address}{Dordrecht}, \bibinfo{year}{1975}).

\end{thebibliography}

\end{document}